\documentclass[useAMS,usenatbib]{mn2e}
\usepackage{graphicx}
\usepackage{times}
\usepackage{natbib}
\usepackage{subfigure}
\usepackage{url}
\usepackage{float,lscape}
\usepackage{amsmath}
\usepackage[dvipsnames]{xcolor}
\usepackage{amssymb}
\usepackage{multirow}
\usepackage{tikz}
\usepackage[titletoc,page]{appendix}

\title[Non-thermal emission from the jet HH80-81]{Detection of non-thermal emission from the massive protostellar jet HH80-81 at low radio frequencies using GMRT}
\author[Vig et al.]{S. Vig$^{1}$\thanks{E-mail: sarita@iist.ac.in}, V. S. Veena$^{1}$, S. Mandal$^{1}$, A. Tej$^{1}$, S. K. Ghosh$^{2}$
\\\\
$^1$Dept. of Earth and Space Science, Indian Institute of Space science and Technology, Thiruvananthapuram, 695 547, India \\
$^2$National Centre for Radio Astrophysics (NCRA-TIFR), Pune, 411 007, India}

\begin{document}

\date{}

\pagerange{\pageref{firstpage}--\pageref{lastpage}} \pubyear{}

\maketitle

\label{firstpage}


\begin{abstract}
Low radio frequencies are favourable for the identification of emission from non-thermal processes such as synchrotron emission. The massive protostellar jet associated with IRAS~18162--2048 (also known as the HH80-81 system) has been imaged at low radio frequencies: 325, 610 and 1300 MHz, using the Giant Metrewave Radio Telescope, India. This is the first instance of detection of non-thermal emission from a massive protostellar jet at such low radio frequencies. The central region displays an elongated structure characteristic of the jet. In addition, the associated Herbig-Haro objects such as HH80, HH81, HH80N, and other condensations along the inner regions of the jet exhibit negative spectral indices. The spectral indices of most condensations are $\sim-0.7$, higher than the value of $-0.3$ determined earlier using high frequency measurements. 
The magnetic field values derived using radio flux densities in the present work, under the assumption of equipartition near minimum energy condition, lie in the range $116-180$~$\mu$G. 
We probe into the hard X-ray nature of a source that has been attributed to HH80, in an attempt to reconcile the non-thermal characteristics of radio and X-ray measurements. The flux densities of condensations at 610 MHz, measured nearly 11 yrs apart,  display variability that could be ascribed to the cooling of  condensations, and emphasize the importance of coeval or nearly-coeval measurements for estimation of spectral indices.
\end{abstract}


\section{Introduction}
\par 

One of the striking phases in the star formation process is the ejection of material at high velocities from the young accreting protostar into the enveloping cloud. 
These jets are believed to disperse angular momentum from the vicinity of the newly forming star to the outer regions and drive large scale outflows seen over the entire stellar mass spectrum.
In addition, jets and outflows are suspected to drive turbulence in molecular clouds, as well as prompt star formation in the vicinity \citep{2007prpl.conf..245A}. In case of massive protostellar systems that are formed deeply embedded in molecular clouds, they contribute to the feedback mechanism that can either halt or induce the infall process \citep{2014prpl.conf..451F}. Numerical simulations predict  the generation of jets from disk winds with the magnetic field playing a prominent role  \citep{2007A&A...469..811Z, 2006ApJ...651..272F, 1997ApJ...482..712O}. The regions close to the launch of protostellar jets have been examined using high resolution spectroscopy in optical wavebands leading to a better understanding of the collimation properties  and velocity structure at the base \citep{2008ApJ...689.1112C}.

\begin{table*}
\begin{center}
\caption{Details of the GMRT observations.}
\label{GMRT}

\begin{tabular}{l c c c c}
\hline  
    Frequency (MHz) &325 & \multicolumn{2}{c}{610} & 1300$^*$\\
 \hline 
 Observation date &2016 Feb 21 & 2005 Jun 20 & 2016 Oct 28& 2016 Feb 27, Mar 27 \\
 On source time (hrs) &7.1 & 5.9 &4.0 & 10.6 \\
 Bandwidth (MHz) &32 &16 &32 & 32\\
 Primary Beam &$86\arcmin.0$ & \multicolumn{2}{c}{$45\arcmin.8$} & $21\arcmin.5$\\
 Synthesized beam &$12\arcsec.9\times 9\arcsec.5$ & $6\arcsec.8\times 6\arcsec.5$ &$6\arcsec.2\times 4\arcsec.3$& $7\arcsec.3\times5\arcsec.0$\\
 Position angle ($ ^\circ $) &46.3 &$-81.8$&18.7 & 54.7\\
 Noise ($\mu$Jy/beam) &190 & 75 &60 & 50\\
 \hline 
 \end{tabular}

\scriptsize{$^*$ Image characteristics quoted for the map produced using the combined data of two days.}\\
\end{center}
 \end{table*}

\par The investigation of jets have been far fewer  in the radio regime as compared to the optical and near-infrared, as the radio emission is usually weak, of the order of few mJy. The radio emission is believed to originate either  (i) from partially ionized jets close to the exciting source (typically within $10-100$ AU) associated with material ejected from the protostar \citep{1996ASPC...93....3A}, or (ii) from shock fronts at the edges of outflows when highly collimated molecular gas at supersonic velocity impinges dense material in its path leading to the formation of Herbig-Haro (HH) objects \citep{1989RMxAA..18...45R}. The shocked medium in the Herbig-Haro objects can have temperatures of millions of degrees leading to emission of X-rays as well \citep{2004A&A...424L...1B,2007A&A...476.1289A,2010A&A...511A...8B}. Most radio detections have been for low-mass protostellar systems with Herbig-Haro objects in our vicinity \citep{2002RMxAA..38..169G}.  A tentative detection of non-thermal emission at low frequencies has been observed for the low-mass protostellar jet of DG-Tau \citep{2014ApJ...792L..18A}. In the case of massive protostars, only a handful
of radio jets have been detected \citep{2011IAUS..275..367R,2016MNRAS.460.1039P} as massive stars are rare, evolve faster, and typically lie at larger distances. All the earlier observations of massive protostellar radio jets have been at higher frequencies ($> 1.4$ GHz) with signatures of non-thermal emission discerned in few cases \citep{1993ApJ...415..191C,1993ApJ...416..208M,1996ApJ...459..193G,1999ApJ...513..775W,2016ApJ...818...27R,2016MNRAS.460.1039P}. While synchrotron emission from a protostellar jet has been detected conclusively solely in one case with the measurement of polarised emission \citep{2010Sci...330.1209C}, the presence of non-thermal emission in other jet systems is usually deduced through the radio spectral indices. If there is a combination of non-thermal and optically thick thermal emission in a region (with the latter anticipated for low radio frequencies), the radio spectral indices are expected to flatten out depending on the contribution of each process, since the slopes have opposite signs for the two processes \citep{2016ApJ...818...27R}. Regions with spectral index larger than -0.1 are attributed to thermal emission \citep{1975A&A....39..217O} while spectral indices $<-0.5$ are believed to be due to non-thermal mechanisms \citep{1999ApJ...527..154K}. Spectral indices at lower frequencies can play a significant role in revealing non-thermal emission \citep{2016MNRAS.456.2425V,2016AJ....152..146N} as the contribution from thermal free-free processes in the vicinity (surmised to be optically thick) is expected to be negligible at these frequencies. 
With this motivation, we have attempted to image the massive protostellar jet associated with IRAS 18162--2048 and estimate the radio spectral indices at low frequencies using the Giant Metrewave Radio Telescope (GMRT).  

\par IRAS 18162--2048 (or GGD 27-28) is a massive protostar at a distance of 1.7~kpc whose luminosity of $2\times10^4$~L$_\odot$ is equivalent to a B0 zero-age main-sequence star \citep{2003ApJ...597..414G}. It possesses an extraordinarily well-collimated parsec scale jet that is believed to arise from an accretion disk associated with the protostar \citep{2012ApJ...752L..29C}. The jet is highly energetic and the measured velocities of condensations lie in the range $600-1400$~km~s$^{-1}$ \citep{1995ApJ...449..184M,1998AJ....116.1940H}. The radio emission from the jet as well as the HH
objects have been the targets of several studies \citep{{1989RMxAA..17...59R},1993ApJ...416..208M,2012ApJ...758L..10M}. \citet{1993ApJ...416..208M} carried out a careful analysis of radio emission in this region using the VLA (20, 6, 3.6 and 2 cm) and find that the knots have spectral indices of $-0.3$ suggesting the non-thermal nature of emission. The jet comprises more than a dozen knots of radio emission with spacing of 1400 AU and widths of less than 500 AU. More recently, \citet{2012ApJ...758L..10M} detected a radio knot further north that is likely to be associated with the jet, thus increasing the size of the jet to 18.7 pc. In addition, polarised radio emission from the radio knots on the jet at $\sim0.5$~pc from the driving source has been discovered, a first for a jet associated with a protostar. This is interpreted as arising due to synchrotron emission from a magnetic field estimated to be $\sim0.2$~mG, that is parallel to the jet axis \citep{2010Sci...330.1209C}. Shocked gas close to the central protostar and the associated HH objects HH80-81 have also been detected in soft X-rays \citep{2004ApJ...605..259P}. In addition, hard X-ray emission from HH80 has been reported by \citet{2013ApJ...776L..22L} who interpret the soft X-rays as arising due to thermal processes and attribute the hard component to synchrotron emission from the shock front.

\par In this paper, we probe the low frequency radio (325 - 1300 MHz) nature of the jet as non-thermal synchrotron emission is expected to be larger at such low frequencies. This is the first case of detection of a radio jet from a massive protostar at these frequencies. The paper is organized as follows. The details of observations and data reduction are given in Section 2. In Section 3, we present our results and analyse them in Section 4. We conclude with a brief summary in Section 5.

\section{GMRT Observations and Data Reduction}
We have mapped the ionized gas emission from IRAS 18162--2048 using the Giant Metrewave Radio Telescope (GMRT), India, in three frequency bands: 325, 610 and 1300~MHz. GMRT comprises of 30 antennas, each of diameter 45~m. The antennas are arranged in a Y-shaped configuration \citep{1991CuSc...60...95S}. Of these, twelve antennas are located randomly within a central region of area $1\times1$~km$^2$ and the remaining eighteen antennas are placed along three arms, each of length 14~km. The shortest and longest baselines are 105~m and 25~km respectively. This type of array configuration enables us to map large and small scale structures simultaneously. The angular sizes of the largest structure observable with GMRT are 32$\arcmin$, 17$\arcmin$ and 7$\arcmin$ at 325, 610 and 1300~MHz, respectively. The first set of observations were carried out at 610~MHz during June, 2005. We then 
followed it up with observations at other frequency bands: 325 and 1300~MHz during February and March, 2016. Considering that the 610 MHz observations were carried out more than a decade earlier compared to the other frequency bands, we observed this region again at 610~MHz in 2016 to check for the variability in flux densities, if any. The bandwidth of observations was 32 MHz in all the frequency bands. The radio sources 3C286 and 3C48 were used as the primary flux calibrators whereas 1822-096 and 1911-201 were used as phase calibrators. The details of the observations are listed in Table~\ref{GMRT}. 

\begin{figure}
\hspace*{-1cm}
\centering
\includegraphics[scale=0.5]{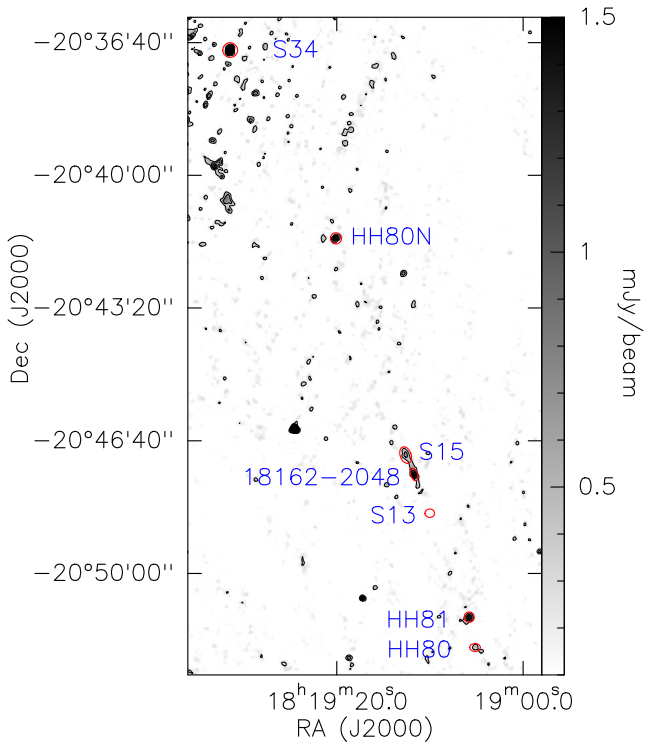}
\caption{Radio contour map of IRAS 18162--2048 region at 610 MHz (2005). The contour levels are at 0.3 mJy/beam to 10 mJy/beam in steps of 0.2 mJy/beam. The Herbig-Haro objects and other jet condensations are shown. The red ellipses denote regions used for estimating flux densities of the condensations. }
\label{full}
\end{figure}

\par The data reduction was carried out using the NRAO Astronomical Image Processing System (AIPS). The data sets were carefully checked and data corrupted due to radio frequency interference, non-working antennas, bad baselines etc. were flagged using tasks such as  $\tt{UVPLT}$,  $\tt{TVFLG}$ and $\tt{UVFLG}$. After the final calibration, the data sets were cleaned and deconvolved using the task $\tt{IMAGR}$. Several iterations of self-calibration were applied to minimize the phase errors. There are two sets of observations for 1300~MHz: one taken on Feb 27 and the other one carried out on Mar 27, 2016. Both these data sets were flagged and calibrated separately and then combined using the task $\tt{DBCON}$. A single continuum map was constructed to increase the signal-to-noise ratio as the earlier observations had 5 non-working antennas. The highest resolution achievable at 1300 MHz with the full UV coverage is $5.6\arcsec\times2.4\arcsec$. Such a high resolution image has a  propensity to display compact and fragmented emission and the elongated jet features are not noticeably revealed.  We, therefore, constructed a low resolution map of 1300~MHz by limiting the UV range to 30~k$\lambda$, giving a beam size of $7.3''\times5.0''$. This serves the dual purpose of (i) identifying the diffuse emission associated with the radio jet at 1300 MHz and (ii) enabling a comparison with the 610 MHz images, on similar spatial scales.

\par As our region of interest is located close to the Galactic plane, it is essential to correct for the contribution of Galactic plane emission to the antenna system temperature (T$_{\textrm{sys}}$) that is significant at low frequencies of 325 and 610~MHz \citep{{2016MNRAS.456.2425V},{2014MNRAS.440.3078V}}. This is because the sky temperature is an integral part of T$_{\textrm{sys}}$ that is estimated using the flux calibrator, located away from the Galactic plane. A correction factor (T$_{\textrm{gal}}$+T$_{\textrm{sys}}$)/T$_{\textrm{sys}}$ is used to scale the flux densities at these frequencies where T$_{\textrm{sys}}$ corresponds to the system temperature associated with the flux calibrators. T$_{\textrm{gal}}$ is estimated by extrapolating the sky temperature value at 408~MHz \citep{1982A&AS...47....1H} to 325 and 610~MHz assuming a spectral index of $-2.6$ \citep{{1999A&AS..137....7R},{2011A&A...525A.138G}} for the Galactic plane emission. 
The correction factors estimated this way are used to scale the images. The flux-scaled images were corrected for the GMRT primary beams to produce the final images. At 1300 MHz, a bright source is observed ($\alpha_{J2000}$: $18^h19^m36.9^s$, $\delta_{J2000}$: $-20^\circ$36$\arcmin$30.7$\arcsec$) that contaminates the flux density measurement of the  radio knot S34 \citep[Source 34, nomenclature from][]{2012ApJ...758L..10M}. The flux density in presence of the contamination due to side lobes from the bright source is 73~$\mu$Jy. In order to remove the effects of the contamination, we have subtracted the cleaned components of this source from the visibility data set, corresponding to a flux density of 181~mJy. Even after the subtraction, we could not observe a considerable variation in the measured flux density of S34 ($<10\%$) leading us to believe that the cleaned components were not well-modeled. Hence, we do not consider the flux density of S34 at 1300~MHz in the estimation of spectral indices.

\begin{table*}
\begin{center}
\caption{Positions, source sizes and  flux densities of the central region and other condensations.}
\label{GMRT_flux}

\begin{tabular}{l c c c c c c c}
\hline  
Source& $\rm{\alpha_{J2000}}$ & $\rm{\delta_{J2000}}$&$\theta_{src}$&325 MHz &\multicolumn{2}{c}{610 MHz} &1300 MHz \\ \cline{6-7}
& $(^{h~m~s})$ &$(^{\degr~\arcmin~\arcsec})$&($\arcsec$)&F(mJy)&F(mJy)$_{2005}$&F(mJy)$_{2016}$ & F(mJy) \\
\hline
HH80&18:19:06.18&$-20$:51:51.69&16.3&$1.17\pm0.52$&$0.95\pm0.27$ & $0.62\pm0.23$& $0.48\pm0.18$ \\
HH81&18:19:06.71&$-20$:51:05.85&15.4&$1.39\pm0.60$&$2.96\pm0.37$ & $1.17\pm0.27$&$0.97\pm0.21$  \\
S13$^*$&18:19:10.77&$-20$:48:31.53&13.7&$1.74\pm0.61$&$0.44\pm0.31$&  $0.94\pm0.278$& $0.70\pm0.21$ \\
18162--2048&18:19:12.11&$-20$:47:30.72&14.4&$2.48\pm0.53$&$3.07\pm0.34$ &  $2.50\pm0.28$&$2.54\pm0.25$  \\
S15$^*$&18:19:12.97&$-20$:47:03.69&20.6&$1.76\pm0.54$& $1.96\pm0.31$ & $1.21\pm0.25$&$0.92\pm0.19$ \\
HH80N&18:19:19.76&$-20$:41:35.25&16.8&$4.43\pm0.71$&$2.52\pm0.38$ & $3.06\pm0.35$&$1.73\pm0.25$ \\
S34$^*$&18:19:30.60&$-20$:36:54.70&22.2&$8.33\pm0.85$&$6.64\pm0.57$ &$4.69\pm0.43$&- \\     
 \hline 
 \end{tabular}
\\
 \scriptsize{$^*$Source nomenclature and flux from \citet{2012ApJ...758L..10M}}\\
\scriptsize{$^a$\citet{1993ApJ...416..208M}}\\

\end{center}
 \end{table*}

\section{Results}

\subsection{HH objects and radio condensations}

\par
\begin{figure*}
\hspace*{-1cm}
\centering
\includegraphics[scale=0.6]{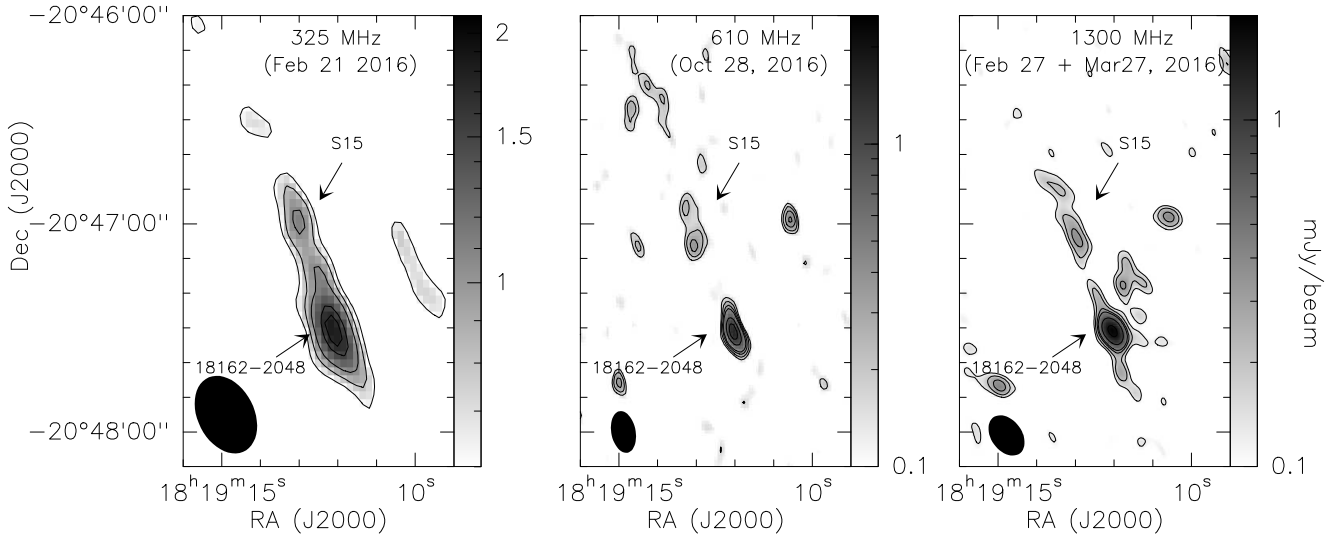}
\caption{Radio maps of IRAS 18162--2048 region at 325, 610 and 1300~MHz (2016 observations). The contour levels are at the following intensity levels: 
325 MHz - 0.6, 0.8, 1.05, 1.5, 1.7 mJy/beam (beam: $12.9''\times9.5''$),
          610 MHz - 0.15, 0.22, 0.32, 0.42, 0.82, 1.2 mJy/beam (beam: $6.2''\times4.3''$), and
          1300 MHz - 0.12, 0.2, 0.32, 0.42, 0.82, 1.6 mJy/beam (beam: $7.3''\times 5.0''$).
The beams are shown towards the bottom left of individual panels.}
\label{gmrt_2016}
\end{figure*}

\begin{figure*}
\hspace*{-1cm}
\centering
\includegraphics[scale=0.6]{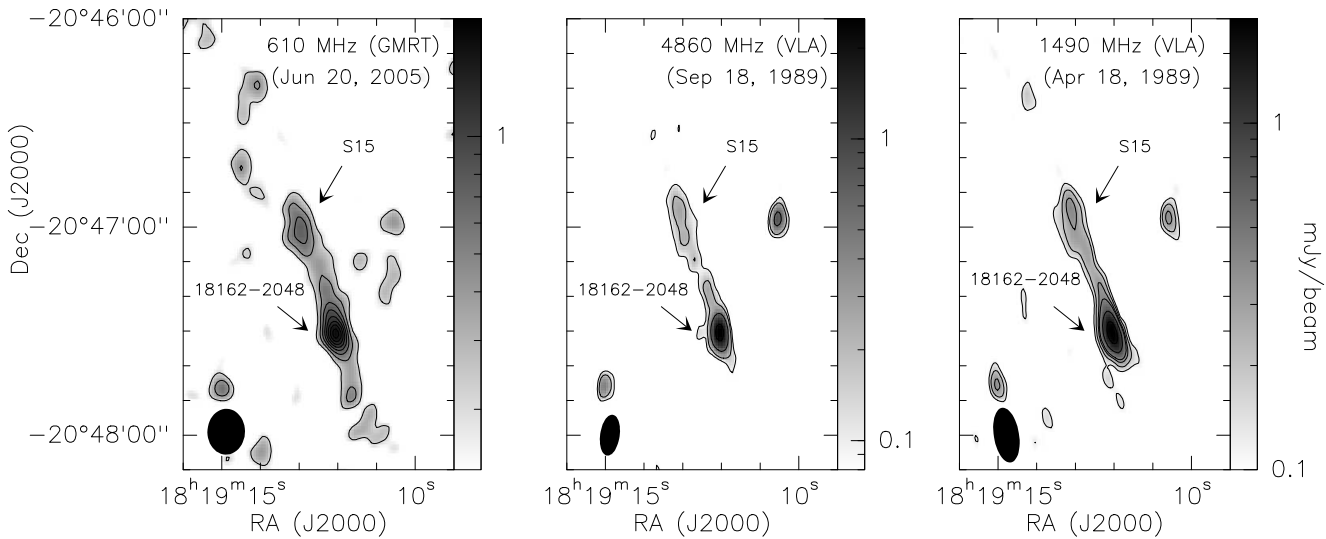}
\caption{Radio maps of IRAS 18162--2048 region at 610 (GMRT), 4860 and 1490 MHz (VLA). The contour levels are at the following intensity levels: 
610 MHz - 0.2 mJy/beam to 1.9 mJy/beam in steps of 0.2 mJy/beam (beam: $6.8''\times6.5''$),
          4860 MHz - 0.09, 0.19, 0.59, 1.9, 3.0 mJy/beam (beam: $5.9''\times3.6''$), and
          1490 MHz - 0.12, 0.2, 0.32, 0.42, 0.82, 1.6 mJy/beam ($8.1''\times4.6''$). The beams are shown towards the bottom left of individual panels.}
\label{610_vla}
\end{figure*}

 The radio jet  associated with IRAS 18162--2048 has been detected at all the three low frequency bands. The ionized emission from the full region at 610 MHz, including the HH objects: HH80, HH81 and HH80N is shown in Fig.~\ref{full}. The central region displaying the inner protostellar jet in 2016 is shown in Fig.~\ref{gmrt_2016}. Fig.~\ref{610_vla} shows the images of the inner jet at earlier epochs: 610 MHz (2005) from GMRT as well as the  4860 and 1490 MHz observations taken with VLA in 1989 \citep{{1993ApJ...416..208M}}. The 4860~MHz image presented in the figure has been extracted from the VLA Archive as part of NRAO - VLA Archive Survey (date of observation: 18 Sept 1989, same as Marti et al. 1993) while the 1490~MHz image has been provided by the Marti et al. group.
In the present work, we focus on the central exciting source, the inner jet condensations as well as the Herbig-Haro objects HH80, HH81, HH80N and Source 34 (referred to as S34 in the present work) . 

\par We have estimated the flux densities of the HH knots as well as other condensations of our interest by selecting elliptical apertures, shown in Fig.~1. The size of the elliptical apertures are selected by visual inspection in all the three bands. The flux densities at all the three frequencies are listed in Table~\ref{GMRT_flux}.  The errors in the flux densities listed in Table~\ref{GMRT_flux} are estimated using the expression $\sqrt{(2\sigma\sqrt{\theta_{src}/\theta_{bm}})^2+(2\sigma')^2}$ where $\sigma$ is the rms noise level of the map,  $\sigma'$ is the the error in flux scale calibration, $\theta_{bm}$ represents the size of the beam, and $\theta_{src}$ is the source size, taken as the geometric mean of the major and minor axes of the elliptical sperture \citep{2013ApJ...766..114S}. The uncertainty in the flux calibration of GMRT is taken to be 5\%  \citep{2007MNRAS.374.1085L}. 

\par As we have two observations of the region at 610 MHz spaced 11 years apart, we compared the flux densities of the central region as well as the outer condensations. These are also presented in Table~\ref{GMRT_flux} and we discern that  all the condensations and the central region display a variability in radio flux at 610 MHz. In order to substantiate the variability of flux densities of the jet condensations,  we have considered four field sources within the primary beam of the 610 MHz images and determined their flux densities at the two epochs: 2005 and 2016, using circular apertures.  The sources and flux densities are listed in Table~\ref{field_flux}. The flux densities of these sources are consistent within 10\%, and the minimal variation could be attributed to calibration procedures. On the other hand, the variation in the jet condensations is between $21-113\%$. All the radio condensations, other than S13 and HH80N, exhibit a decrease in flux densities from 2005 to 2016. We speculate on the possible reasons in Sect. 4.3. A comparison of flux densities at 20~cm reported in this work, with those published previously \citep[Table 2 of][]{1993ApJ...416..208M} implies a lowering of values  by factors between $2-3$. We are of the view that variability has an important role to play in this decrease although the slightly variant beam sizes could also contribute to the difference.

\par We next examine the morphology of the inner jet as a function of time. We search for changes by considering the two segregated group of images: (i) Inner jet in 2016 at various frequencies shown in Fig.~\ref{gmrt_2016}, and (ii) Inner jet prior to 2016, evident from the Fig.~\ref{610_vla}. The latter group includes the 610 MHz image obtained in 2005 as well as VLA images  at 1490 and 4860~MHz \citep[Project code AR209;][]{1993ApJ...416..208M}. 
A visual scrutiny of S15 suggests that while it has been identified as a single source from the VLA data, this source displays sub components in the recent observations at 610 as well as 1300~MHz. The resolution of the 1490 and 4860~MHz maps are 6$\arcsec$ and  5$\arcsec$, respectively; similar to the beam-sizes of 610 and 1300~MHz images ($\sim$6$\arcsec$). Hence, this effect is unlikely to be due to resolution effects. Rather, we suspect this apparent transformation is due to the evolution of the condensations as they advance in the ambient medium. However, as the signal-to-noise ratio is poor, we refrain from a detailed characterisation of the knot components and their flux densities.

\begin{table}
\begin{center}
\caption{Flux densities at 610~MHz (using 2005 and 2016 observations) of 4 random sources in this field.}
\label{field_flux}

\begin{tabular}{l c c c c}
\hline  
Source& $\rm{\alpha_{J2000}}$ & $\rm{\delta_{J2000}}$&\multicolumn{2}{c}{610~MHz} \\ \cline{4-5}
& $(^{h~m~s})$ &$(^{\degr~\arcmin~\arcsec})$&F(mJy)$_{2005}$&F(mJy)$_{2016}$ \\
\hline
1 & 18:19:00.61 & $-20$:58:06.15 & $1.50\pm0.32$ & $1.47\pm0.21$ \\
2 & 18:19:24.84 & $-20$:46:21.52 & $9.03\pm0.95$ &  $9.20\pm0.91$ \\
3 & 18:18:30.18 & $-20$:55:34.16 & $0.61\pm0.23$ & $0.65\pm0.22$ \\
4 & 18:19:37.06  & $-20$:49:27.17 & $1.68\pm0.34$ & $1.84\pm0.19$ \\

 \hline 
 \end{tabular}
\\
\end{center}
 \end{table}
\subsection{Spectral indices}

 The spectral indices of the HH knots and condensations have been estimated using flux densities measured at all the low frequency bands. The spectral index $\alpha$ is defined as  $F_\nu\propto$ $\nu^{\alpha}$, where $F_{\nu}$ is the flux density at frequency $\nu$. The spectral indices are determined using the flux densities listed in Table~\ref{GMRT_flux} and shown in Fig.~\ref{specin}. We prefer to use the flux densities directly rather than create the spectral index maps as (i) the flux densities of the inner jet and condensations are relatively low, of the order of few mJy or lower, and (ii) the radio emission is over scales of the beam or marginally larger. 
The variability of flux densities observed at 610 MHz highlights the importance of nearly simultaneous measurements to determine the spectral indices.  We have, therefore, used the flux densities at 610 MHz from observations carried out in 2016 to evaluate the spectral indices. Although these observations were taken nearly six months after the observations in the other bands, we proceed under the assumption that flux densities have not changed appreciably.  The spectral indices are listed in Table~\ref{mag}. All the sources except the central exciting source, 18162--2048, display negative spectral indices, $\alpha_\textrm{GMRT}<-0.2$. This accentuates the non-thermal contribution to the emission. The spectral index of the central knot associated with the exciting source of 18162--2048 is nearly flat,  $\alpha_\textrm{GMRT}\sim+0.02\pm0.01$. At such low frequencies, the thermal free-free emission is expected to be optically thick ($\alpha\sim2$). The flatness of the spectral index can be interpreted as a combination of thermal emission from the ultracompact HII region associated with the exciting source and non-thermal emission from the jet in its vicinity. This reasoning corroborates alternate high resolution observations of the jet where condensations have been perceived, that lie symmetrically on either side of the central source within $1''$, along the direction of the jet \citep{1995ApJ...449..184M}. 
The inner condensations identified as S13 and S15 display spectral indices of $-0.7\pm0.2$ and $-0.5\pm0.1$, respectively, while HH80 and 81 exhibit spectral indices of $\alpha_\textrm{GMRT}\sim -0.6\pm0.2$ and $-0.3\pm0.1$, respectively. The knot HH80N has a spectral index of $-0.7\pm0.2$ while S34 displays the steepest spectral index ($-0.9$) among the condensations considered. For S34, we have only considered the lower two frequency bands as the flux density at 1300~MHz is vitiated by the bright source in its proximity (Sect. 2). The spectral indices of condensations other than HH81 are steeper than the previously reported values using higher frequency VLA measurements (see Table~\ref{mag} for a comparison).

\begin{table*}
\begin{center}
\caption{Spectral indices and other properties of the central region and the condensations.}
\label{mag}

\begin{tabular}{l c c c c c}
\hline  
Source& Spectral Index   &Energy Power-law exponent$^a$ & $B_{\rm eq}^b$ & $u_{\rm part}^c$ & Spectral index\\ 
& ($\alpha_\textrm{GMRT-2016}$) & ($\delta$) & ($\mu$G) & (10$^{-10}$~erg/cm$^{-3})$ &($\alpha^d_\textrm{VLA}$)\\
\hline
HH80&$-0.63\pm0.19$&$2.3$ & 132 & 9 & $-0.3\pm0.1$ \\
HH81&$-0.26\pm0.01$&$1.8$ & 160 & 14 &$-0.3\pm0.1$ \\
S13$^*$&$-0.65\pm0.17$ &$2.3$ & 152 & 12 & $-0.4\pm0.2$\\
18162--2048&$+0.02\pm0.01$ &- & - & - & $+0.2\pm0.1$\\
S15$^*$&$-0.47\pm0.06$ & $1.9$ & 116 & 7 & $-0.4\pm0.1$\\
HH80N&$-0.68\pm0.04$ &$2.4$ &  180 & 17 & $-0.3\pm0.1$\\
S34$^*$&$-0.9$ & $2.8$ &  174 & 16 &$-0.46\pm0.02$\\     
 \hline 
 \end{tabular}
\\
 \scriptsize{$^*$Source nomenclature from \citet{2012ApJ...758L..10M}}\\
\scriptsize{$^a$ Power law exponent of the assumed electron energy distribution: $N(E)dE\propto E^{-\delta} dE$}\\
\scriptsize{$^b$ Magnetic field obtained under assumption of equipartition that approximately corresponds to energy minimisation condition.}\\
\scriptsize{$^c$ Total energy density under minimisation condition.}\\
\scriptsize{$^d$ Spectral indices from high frequency VLA observations of \citet{1993ApJ...416..208M} and \citet{2012ApJ...758L..10M}.}\\

\end{center}
 \end{table*}

\par We observe a general trend of steepening of spectral index of condensations (except HH81) with distance from the central source, 18162--2048. The northern condensations of the jet in order of increasing distance are
S15, HH80N and S34 with spectral indices: $-0.5$, $-0.7$ and $-0.9$, respectively. The southern condensations include S13, HH81 and HH80 with spectral indices: $-0.7$, $-0.3$ and $-0.6$. 
A steepening of spectral indices is expected for the outer condensations if the density of the ISM in which the jet is emitted remains nearly homogeneous. This is because the distant condensations are older and particle acceleration mechanism across shock waves are expected to weaken with time \citep[][]{2017arXiv170807011W}. The northern side of the jet, that corresponds to the blue lobe of the CO outflow moving into the molecular cloud \citep{{1989ApJ...347..894Y},{1993ApJ...416..208M}}, satisfies the general steepening law. 
This trend of spectral indices is not observed in the southern condensations as HH81 has a lower spectral index compared to HH80. The low spectral index of HH81 can plausibly be justified by the compactness of this knot, as a result of which the particles are unable to dissipate their energy. The compact morphology of HH81 has been revealed using high-resolution imaging in optical wavebands by \citet{1998AJ....116.1940H}, which is unlike HH80 that displays intricate and elongated sub-structures. 

\section{Discussion}

\subsection{Magnetic field estimates in the radio condensations}

The non-thermal emission implied through the negative spectral indices in the radio condensations is attributed to synchrotron emission from relativistic particles that are diffusion shock accelerated in the magnetic field associated with the jet. The total synchrotron emission furnishes the strength of the magnetic field while the field uniformity and structure can be extracted from the degree of polarisation of emission. The polarisation measurements by \cite{2010Sci...330.1209C} have unveiled the magnetic field that is parallel to the jet and extends upto 0.5~pc from the central exciting source. Here, we estimate the magnetic field strength in the individual radio condensations extending from S34 to HH80. In order to estimate the magnetic field strength, we employ the classical assumption about the minimisation of the total energy content of the synchrotron source. This approximately corresponds to the equipartition of energy between the magnetic field and the relativistic  particles. This formalism is considered for the standard energy distribution of electrons of the form: $N(E)dE\propto E^{-\delta} dE$ where $N(E)$ represents the number of electrons having energy between $E$ and $E+dE$ and $\delta = 1-2\alpha$, presumed to be homogeneous across the condensation. For the region of synchrotron emission delineated by an ellipse of size $\theta_x$ and $\theta_y$, the magnetic field under minimum energy  condition ($ B_{\rm eq}$) is given by the following expression \citep{1980ARA&A..18..165M}: 
$$ \left(\frac{B_{\rm eq}}{\rm gauss}\right) = 5.69\times10^{-5}\left[ \frac{1+K}{\eta(\sin{\phi})^{3/2} (\alpha + 1/2)}  \left( \frac{\rm arcsec^2}{\theta_{\rm x} \theta_{\rm y}}\right) \right]^{2/7} $$
\begin{equation}
  \times \left[\left( \frac{\rm kpc}{s}\right) \left( \frac{F_{\rm o}}{\rm Jy}\right)  \frac{(\nu_2^{\alpha+1/2} - \nu_1^{\alpha+1/2})}{\nu_{\rm o}^\alpha (\alpha+1/2)}\right]^{2/7} 
\end{equation}

\noindent Here, $K$ is the ratio of energy of heavier ions (particularly relativistic protons) to the energy of the relativistic electrons, $\eta$ is the beam filling factor, $\phi$ is the angle between the uniform
magnetic field and the line-of-sight, $s$ is the path length through the source in the line-of-sight, $F_{\rm o}$ is the radio flux density at frequency $\nu_{\rm o}$ and $\alpha$ is the radio spectral index between the cut-off frequencies $\nu_1$, and $\nu_2$. In the above equation, the frequencies, $\nu_{\rm o}$, $\nu_1$ and $\nu_2$ are  expressed in GHz. For estimation of the magnetic field, we assume $K\sim40$ for Fermi accelerated electrons typical of astrophysical environments \citep{2005AN....326..414B}, $\eta\sim0.5$ to account for clumpiness within the condensations that is evident from the detailed structure of HH80 and HH81 objects \citep{1998AJ....116.1940H}, and $\phi\sim34^\circ$ \citep{2012ApJ...758L..10M}. We consider the GMRT flux densities at frequency of $\nu_{\rm o}=0.610$~GHz (2016) while the cut-off frequencies are taken as $\nu_1=0.01$~GHz and $\nu_2=100$~GHz. For each condensation, $s$ is estimated using $\theta_{\rm src}$ whose values are listed in Table~\ref{GMRT_flux}. The spectral indices, magnetic fields and particle energy densities ($u_{\rm part}=B_{\rm eq}^2/6\pi$) derived for the six condensations are listed in Table~\ref{mag}. We find that the magnetic field ranges between $116-180$~$\mu$G.  The power-law exponent of the electron energy distribution, $\delta$, lies in the range $1.8-2.4$ while the particle energy density has values between $1-1.7\times10^{-9}$~erg/cm$^{-3}$. It is important to be cognizant of the fact that Eqn. (1) and, therefore, the estimates of the field and energy densities posses inherent uncertainties. In particular, the uncertainties are related to the values of $K$, $\nu_1$ and $\nu_2$. The value of $K$ is unknown and is expected to lie between 1 and 2000. A value of 100 is considered appropriate for electrons produced following collisions in a circumstellar medium \citep{1970ranp.book.....P} but \citet{1980ARA&A..18..165M} adopted a value of $K=1$ for minimum energy conditions. In this case, if we consider $K=100$, the magnetic field values are higher by $\sim30$\%. The upper and lower cut-off frequencies over which the relativistic particles radiate, i.e $\nu_1$ and $\nu_2$, are also uncertain and the values taken are those that are archetypal for astrophysical jets from literature \citep{{1980ARA&A..18..165M},{2004IJMPD..13.1549G}}. 
Due to the shape of the power-law distribution of electron energy, the lower frequency cut-off limit is significant. In addition, the estimate of magnetic field is revised when one considers a cut-off in energy rather than a cut-off in frequency of the emitted synchrotron spectrum. With an energy cut-off, the exponent of the terms on the right hand side of Eqn. (1) would be revised from 2/7 to 1/($3+\alpha$) \citep{2005AN....326..414B}. Writing the energy of electron in terms of its Lorentz factor and considering $\gamma_{\rm min}\sim100$, the change in the magnetic field estimate is within 10\%. 

\par The equipartition magnetic field estimates obtained towards the jet condensations are consistent with the value of 90~$\mu$G determined towards the northern region by \citet{2016ApJ...824...95K} using the Chandrasekhar-Fermi method that utilises the dispersion of the measured near-infrared polarisation angle. It is also in congruence with the estimate of 0.2 mG by \citet{2010Sci...330.1209C} using the equipartition method. We would like to bring attention to the fact that the equipartition magnetic field estimates in the previous works have considered a radio spectral index value of $-0.3$ which is shallower for condensations other than HH81. \citet{2010A&A...511A...8B} consider the value of magnetic field towards HH80 using models that constrain the observed emission, by two populations of particles: (i) relativistic primary electrons, and (ii) relativistic secondary electrons produced by inelastic proton-proton collisions in the shock that also accelerates the protons. They employ magnetic field values of 3~$\mu$G for primary electrons and 2.5~mG for secondary electrons, under the assumptions listed in Table~2 of their work. 

\subsection{High energy emission by relativistic electrons}

\par In the region associated with IRAS~18162--2048, X-ray emission has been detected from the Herbig-Haro objects HH80-81 as well as from the vicinity of the central source \citep{{2004ApJ...605..259P},{2013ApJ...776L..22L},{2009ApJ...690..850P}}. This is explained on the basis of strong shocks occurring when the southern
extension of this bipolar outflow collides with the ambient material at high velocity.
Towards HH80, \cite{{2013ApJ...776L..22L}} observed both: (i) soft X-ray sources  ($\sim 0.11$~keV) attributed to thermal emission, and (ii) a non-thermal hard X-ray source with flux $\sim 10^{-14} \mathrm{erg\, cm^{-2}\, s^{-1}}$ (between 0.4 and 10~keV). They associate the thermal contribution to the hot post-shock region where the jet slams into the dense ambient medium, whereas the hard component is ascribed to the synchrotron radiation at the shock front. The origin of the hard-component is, however, speculative. Assuming this hard X-ray emission to be associated with HH-80, we are interested in ascertaining the mechanism by which this radiation is produced and to this end, we resort to the non-thermal emission in radio wavebands. 

\par We reconcile the fluxes in the radio and hard X-ray bands by assuming that the same population of relativistic electrons that emit in radio by synchrotron process is also responsible for the hard X-ray emission. We consider various processes of emission by relativistic electrons at hard X-ray energies. 
In the first case, we assume that the hard X-ray spectrum is an extension of the radio synchrotron emission spectrum with the same spectral index.  The hard X-ray flux contribution between energies $E_1$ and $E_2$ due to synchrotron emission with spectral index $\alpha$ from these relativistic electrons can be expressed as 
\begin{equation}
 F_\mathrm{X(E1-E2)}=\frac{F_{\rm R}}{\nu_{\rm R}^{\alpha}} \frac{(E_2^{1+\alpha} - E_1^{1+\alpha})}{(1+\alpha)h^{1+\alpha}}
\end{equation}

\noindent Here $F_{\rm R}$ is the observed radio flux at frequency $\nu_{\rm R}$, and $h$ is Planck's constant. We have considered the 610~MHz flux density of HH80 observed in 2016 for estimating the hard X-ray contribution and taken $\alpha$ to be -0.63 based on the observations. For energies between $E_1=0.4$~keV and $E_2=10$~keV. We evaluate the hard X-ray flux as $2.4 \times 10^{-14} \mathrm{erg\, cm^{-2}\, s^{-1}}$. Although this estimate is comparable to the observed hard X-ray flux, we note that the observed flat hard tail \citep{{2013ApJ...776L..22L}} has an energy spectral index of $+0.2$, which is inconsistent with the observed radio spectral index ($-0.63$ for HH80).  

\par The second possibility that we consider is that the hard X-ray flux is due to synchrotron self-Comptonization of the radio photons by the same population of relativistic electrons. For this, we first estimate the radio flux density by considering a population of relativistic electrons with the energy distribution described earlier [$N(E)dE\propto E^{-\delta} dE$] within a finite emitting region, accelerated in a given magnetic field. We take the size of the emitting region to be typically $\sim0.1$~pc (region over which the HH80 radio flux has been integrated), distance to the source $\sim 1.7$~kpc and find that the observed radio flux density (Table \ref{GMRT_flux}) can be explained by assuming a typical interstellar (ISM) magnetic field $B =3~\mu G$ \citep{2006ApJ...642..868H} and a non-thermal electron density $n_{\rm e}=5 \times 10^{-3}\mathrm {cm^{-3}}$. The low value of number density suggests that the large radio emitting region considered (0.1~pc) is not uniform, but rather clumpy. This is anticipated, as the acceleration of electrons to relativistic energies is expected near the shock head and corroborated by the filamentary and clumpy morphology of HH80 in H$\alpha$ by \citet{1998AJ....116.1940H}. Hence, a volume filling factor $f \sim 10^{-3}$ is required to explain the observed radio flux with typical ISM number density of $n_{\rm e}=10$~cm$^{-3}$. However, we note that the value of magnetic field in this case is more than an order of magnitude lower than the equipartition value (132~$\mu$G). Employing the equipartition value for an electron density of $n_{\rm e}=10$~cm$^{-3}$, the radio flux density is consistent with a smaller emitting region of size $\sim200$~AU. 
Assuming the radio emission at the synchrotron peak frequency ($\nu_{\rm R}$) from a relativistic electron, the Lorentz factor of the relativistic electron is $\gamma_{\rm e} \propto \sqrt{\nu_{\rm R}/{\rm B}}$. Taking typical values of $\nu_{\rm R}=1$~GHz and $B = 3$~$\mu$G, the electron Lorentz factor is $\gamma_{\rm e} \sim 9\times 10^3$. We estimate the Compton {\it y-parameter} for Thomson scattering ($y\propto\gamma^{2-\delta}$) considering an average size of emitting region as 0.1 pc, and a number density $10\, \mathrm{cm^{-3}}$ for HH80 complex. We find that $y \sim 10^{-8}$ and the calculated X-ray flux due to synchrotron self-Comptonization in the energy range: $0.4-10$~keV, is $\sim 10^{-21} \mathrm {erg\, cm^{-2}\, s^{-1}}$, that is substantially lower than the observed value. Hence, the effect of synchrotron self-Comptonization is insignificant. 

\par The third alternative for the origin of hard X-ray emission could be the inverse-Compton scattering (IC) of the background thermal infrared photons by the relativistic non-thermal electron population. The IC contribution to the X-ray flux density can be estimated using the following expression, that determines the integrated X-ray flux density, $F_{\rm X(E_1-E_2)}$, between energies $E_{\rm 1}$ and $E_{\rm 2}$ corresponding to a synchrotron flux density $F_{\rm R}$ at the radio frequency $\nu_{\rm R}$ \citep{2004IJMPD..13.1549G}:

$$\left( \frac{F_{\rm X(E1-E2)}}{\rm erg\,s^{-1}\,cm^{-2}}\right)  = g(\alpha) \left( \frac{F_{\rm R}}{\rm Jy}\right)  \left[ 0.0545 \times \left(\frac{\rm MHz}{\nu_{\rm R}}\right)\right]^\alpha \left( \frac{\rm \mu G}{B}\right)^{1-\alpha}$$
\begin{equation}
 \times \left[\left(\frac{E_{\rm 2}}{\rm keV}\right)^{1+\alpha} - \left(\frac{E_{\rm 1}}{\rm keV}\right)^{1+\alpha}\right]\left( \frac{T}{\rm 2.7 K}\right)^{3-\alpha} 
\end{equation}

\noindent Here, $\alpha$ is the radio spectral index and $g(\alpha)$ represents the IC parameter that is a function of spectral index, whose values are listed in Table 2 of \citep{2004IJMPD..13.1549G}. $T$ represents the temperature of the background photons that are inverse-Comptonized by the relativistic particles. As earlier, $E_1=0.4$~keV, $E_2=10$~keV, and  $B=3~\mu \mathrm{G}$ (required to match the radio flux as well). We consider the background photon temperature $T=14$~K for the envelope of the molecular cloud \citep{2011ApJ...738...43M}. Using these values, the IC contribution is found to be $5.9\times10^{-14} \mathrm{erg\,s^{-1}\,cm^{-2}}$ for HH80 which is in close agreement with the observed hard X-ray flux.  However, we note that the hard X-ray flux is $1.0\times10^{-18} \mathrm{erg\,s^{-1}\,cm^{-2}}$  for the equipartition magnetic field, that is nearly four orders of magnitude lower.

\par Thus, among the three models considered, the inverse-Comptonization of the thermal background photons appears to explain the radio and X-ray flux measurements as well as the observed spectral indices. As
the magnetic field value employed in this framework is far from the equipartition value, higher resolution radio measurements of non-thermal emission are required in order to rigorously ascertain the association of the hard X-ray source with HH80.
We note that \cite{2010A&A...511A...8B} argued that the high energy X-ray may be due to emission from the secondary particles accelerated across the shock head. However, for the secondary electrons case, these authors adopt a value of magnetic field that is nearly three orders of magnitude higher (2.5~mG) than the primary electrons (3~$\mu$G).

\subsection{Variability in radio flux densities}

\par
\begin{figure*}
\centering
\hspace{-0.5cm}\quad \includegraphics[scale=0.33]{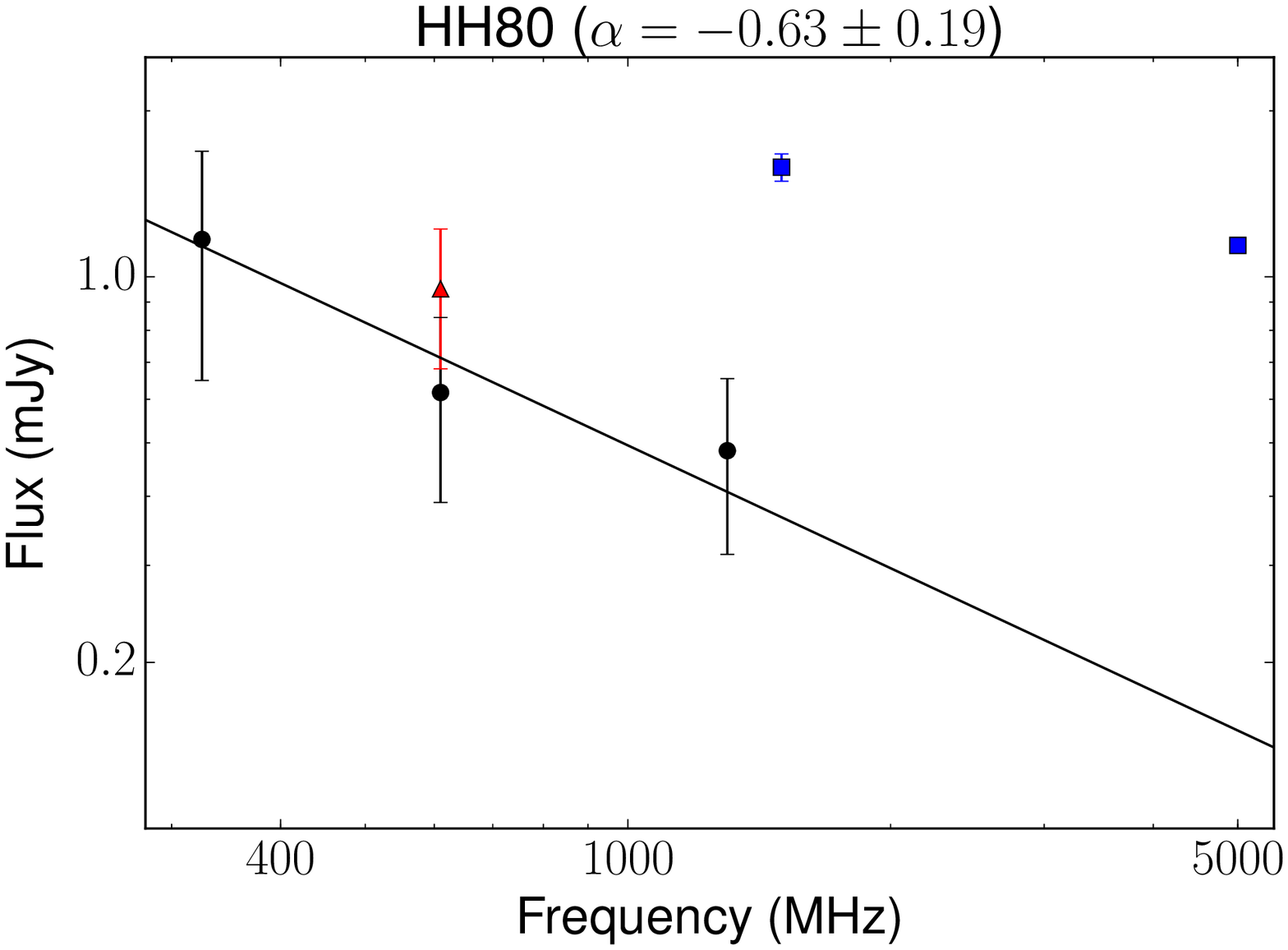}\hspace{-0.5cm} \quad \includegraphics[scale=0.33]{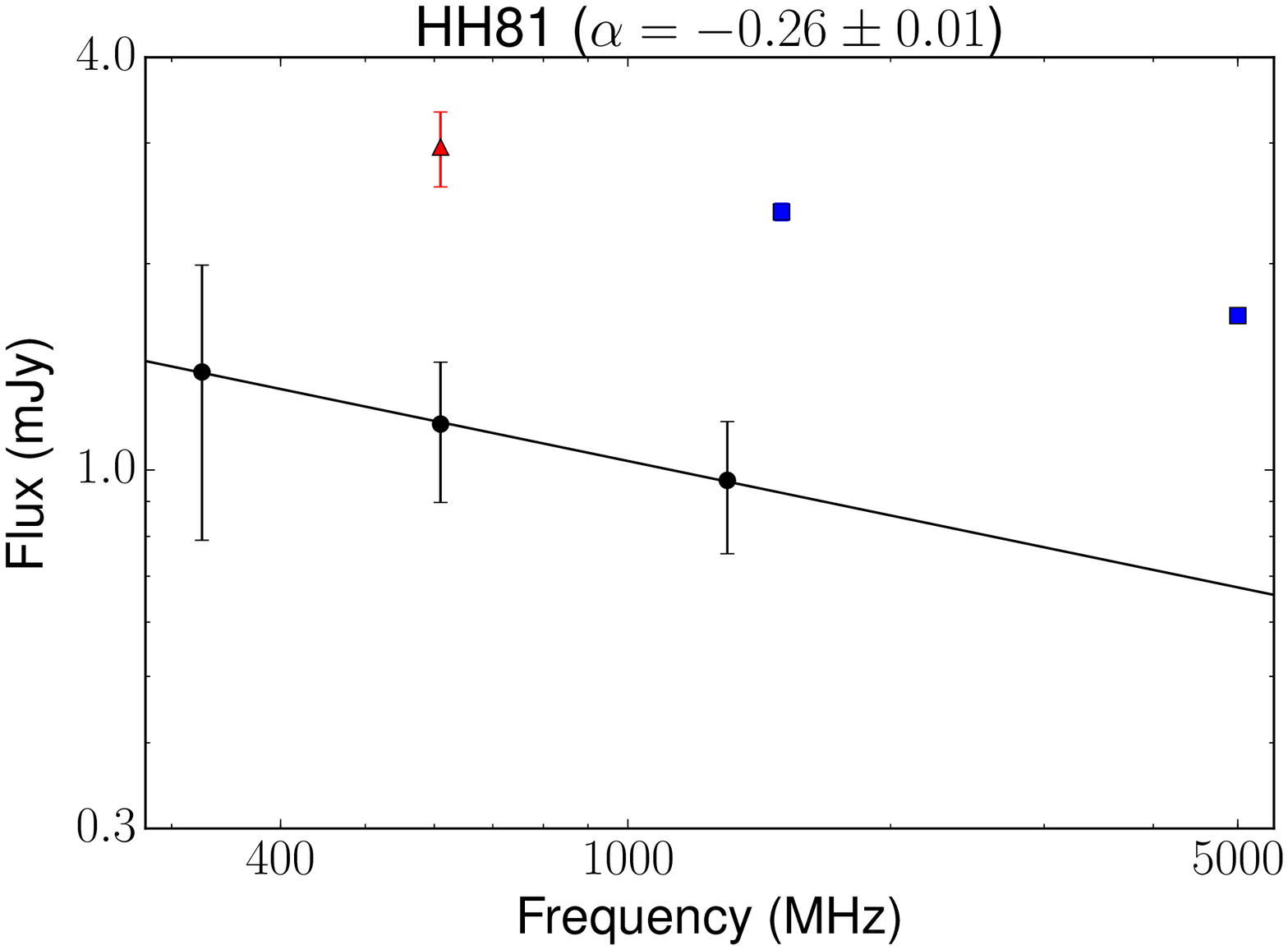} \hspace{-0.5cm}\quad \includegraphics[scale=0.33]{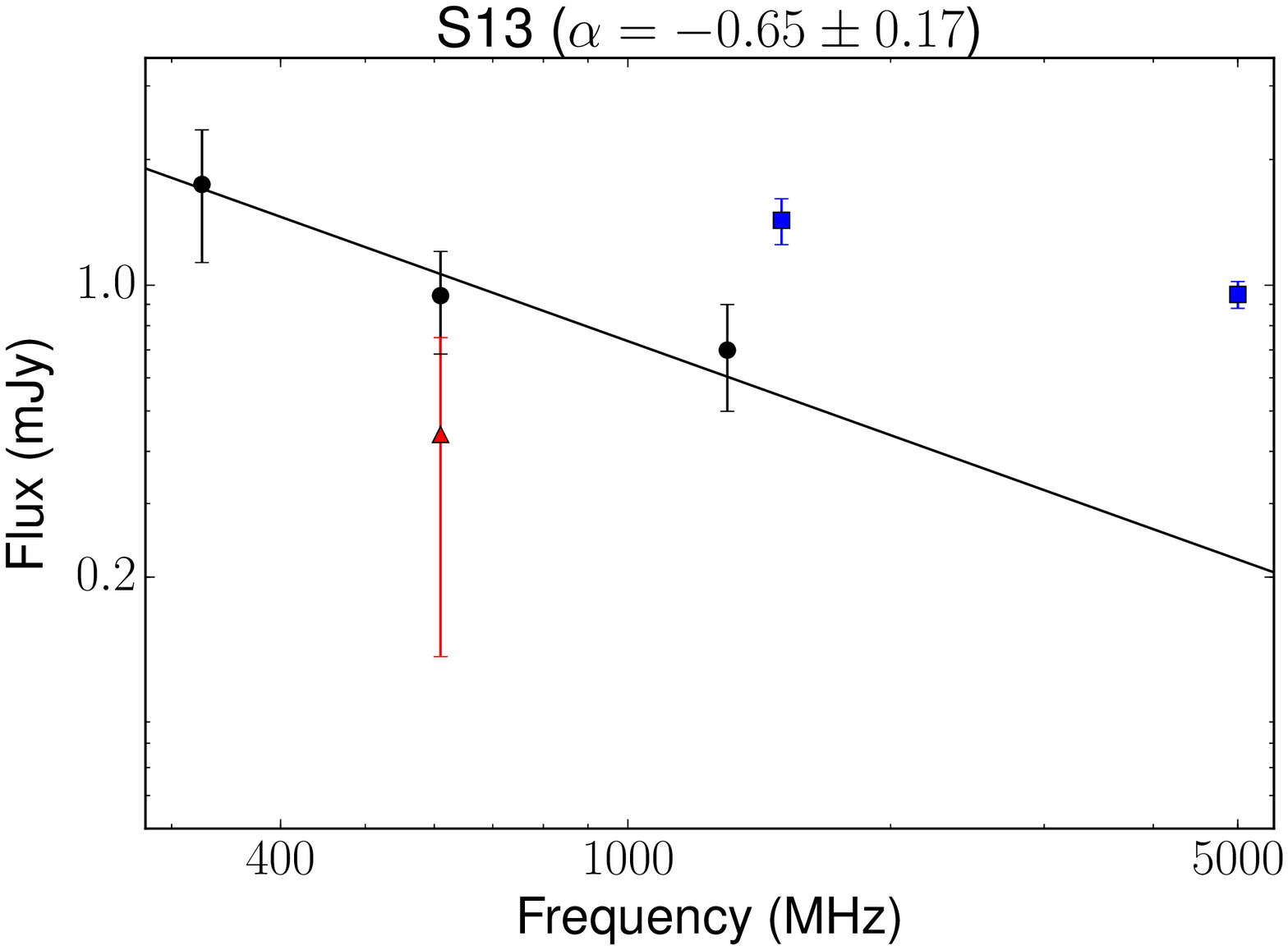} 
\hspace{-0.5cm}\quad \includegraphics[scale=0.33]{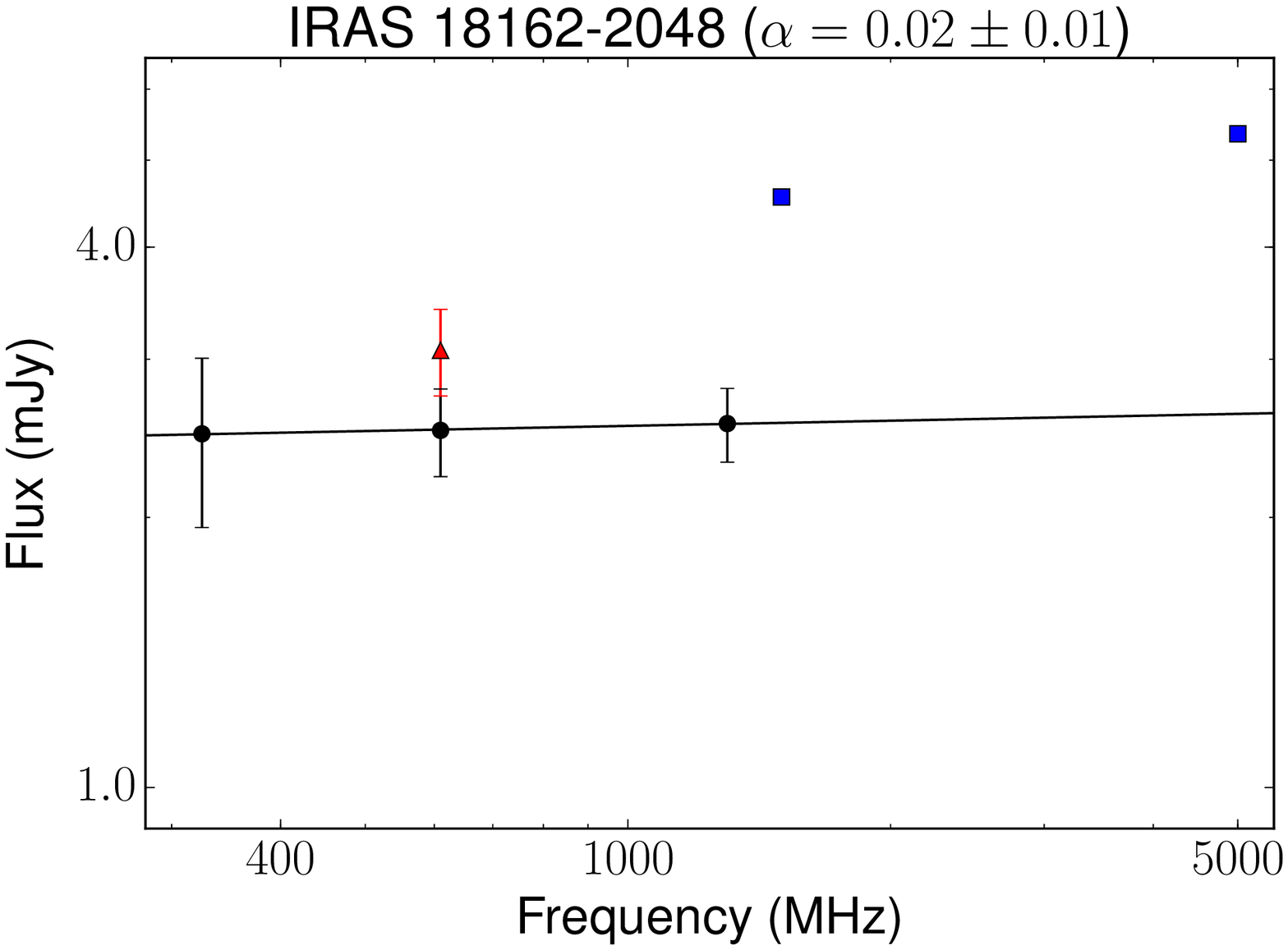}\hspace{-0.5cm} \quad \includegraphics[scale=0.33]{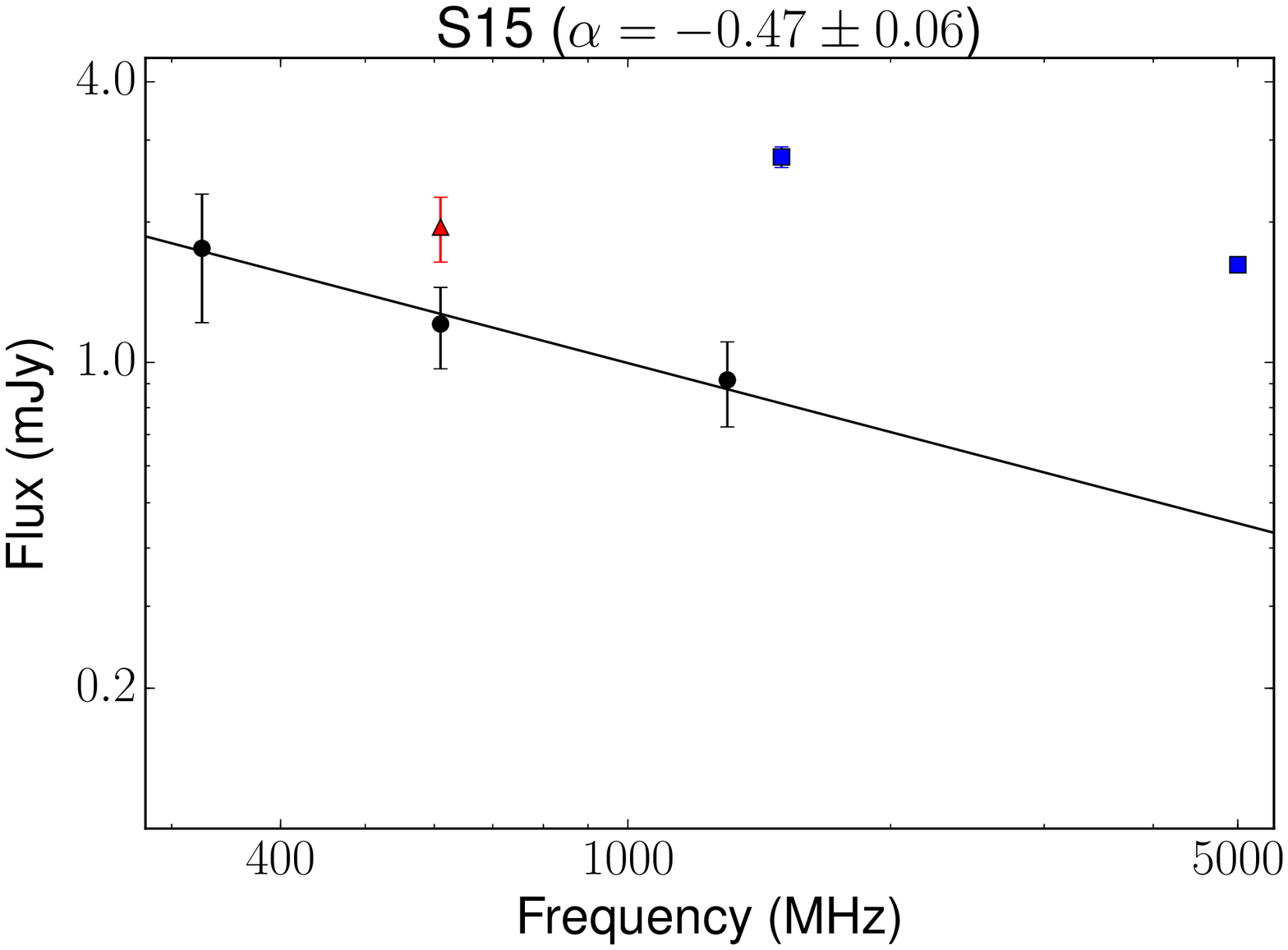} \hspace{-0.5cm}\quad \includegraphics[scale=0.33]{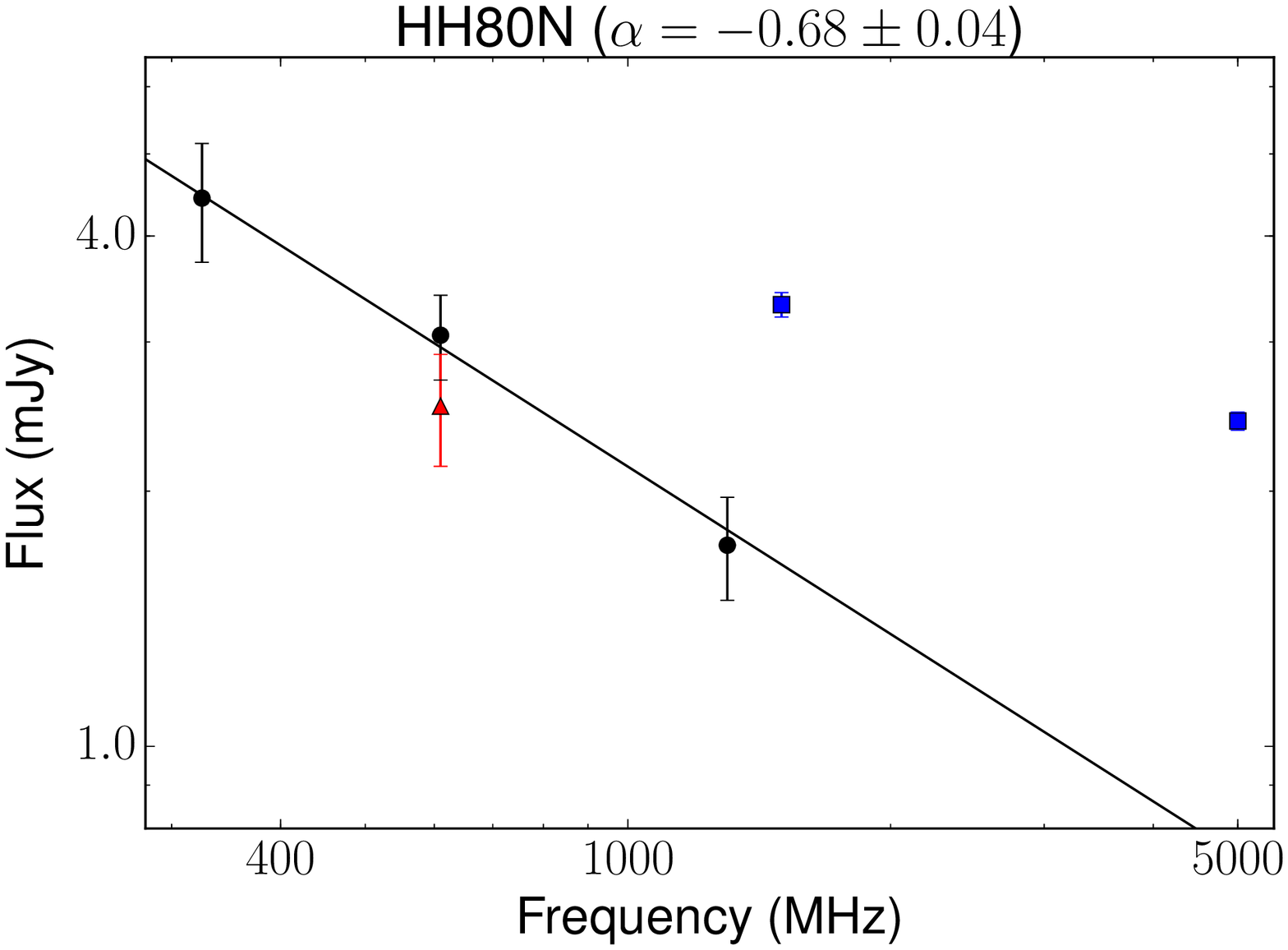}
\hspace{-0.5cm}\quad \includegraphics[scale=0.33]{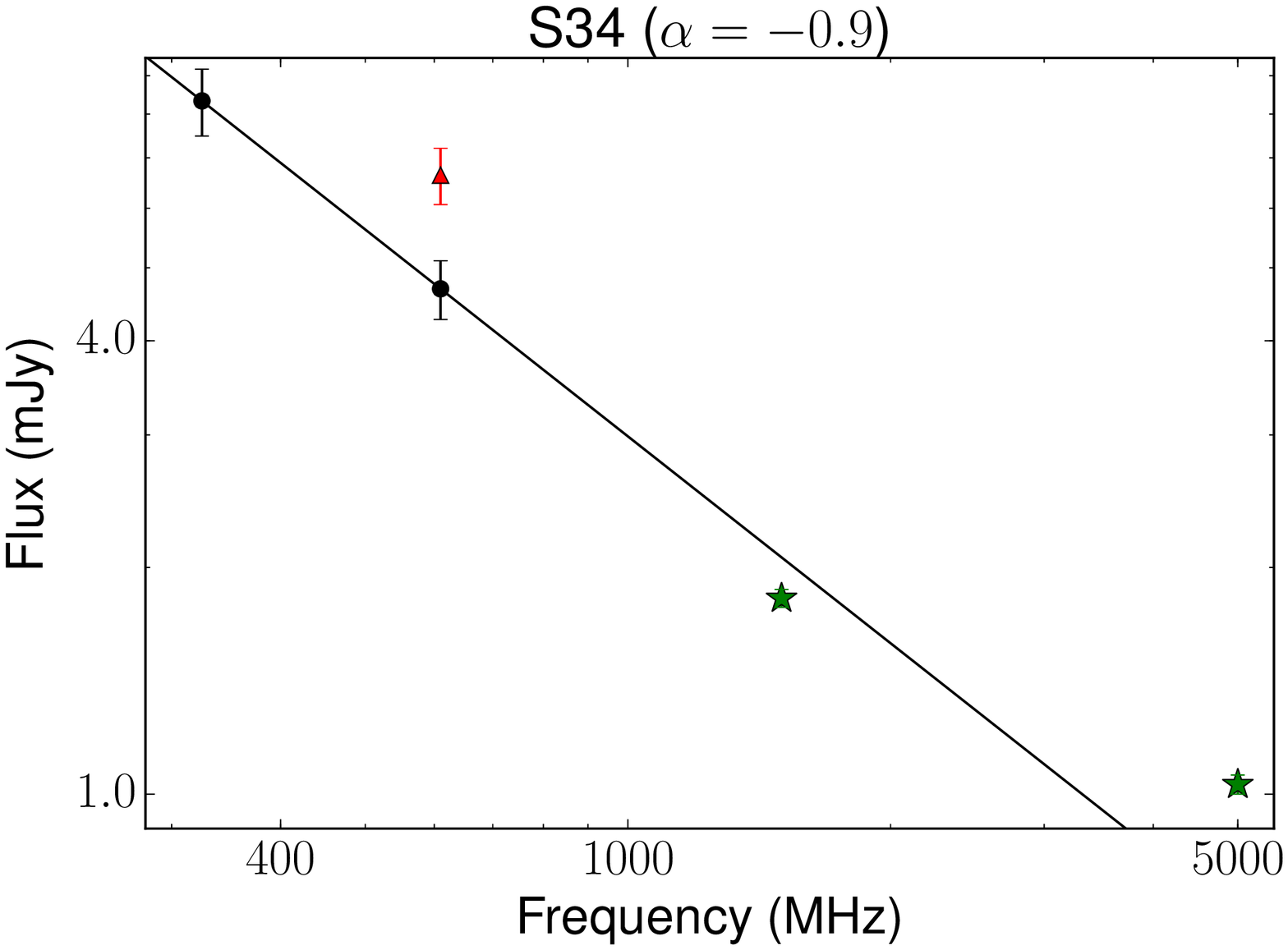} 
\caption{Spectral index plots of the condensations and the central region using flux densities at 325, 610 and 1300~MHz (filled circles) observed in 2016. The triangle in each plot represents the corresponding 610~MHz flux density measured in 2005. The high frequency flux densities from earlier epochs are shown using squares  \citep{1993ApJ...416..208M} and star-symbols \citep{2012ApJ...758L..10M}.}
\label{specin}
\end{figure*}

 
Reports of radio variability from protostellar jets in literature have been sporadic and mostly related to the intrinsic variability of the central exciting sources of the jets \citep{2004RMxAA..40...31G, 2012A&A...537A.123R,2002AJ....124.1045R}. There have been instances where UV, optical and infrared variability in the emission from Herbig-Haro objects have been investigated \citep{1985ApJ...292L..75B,1996A&A...306..255L,2016AJ....151..113R}. Here, we speculate on the likely origin of the radio flux variability under the assumption that non-thermal  processes dominate the radio contribution and refrain from any quantitative analysis as we have solely two measurements at one frequency, spread more than a decade apart. The variability is related to (i) the ambient medium through which the condensations are moving, and (ii) time evolution of the shock. If there is no significant change in the density, then we expect a reduction of radio flux and steepening of spectral index due to cooling mechanisms (including adiabatic expansion and radiative processes) leading to a weakening in the particle acceleration mechanism. We observe a decrease in the 610 MHz radio flux density (black circles in Figure \ref{specin}) in all condensations (except HH80N and S13) when compared to the previous measurements (triangles in Figure \ref{specin}). For S13, the uncertainty in the earlier estimate is significantly large and for HH80N the change in flux density is marginal. A reduction in the ambient density with time as the condensation moves through the medium could accelerate the relativistic electrons further leading to a decrease in spectral index and increase in radio flux. Thus, the variability observed in the radio condensations, along with their proper motions, can be used to probe the interstellar medium through which the condensations advance. This highlights the importance of coeval or nearly simultaneous observations to determine the spectral indices.

\section{Summary}

Low-frequency radio observations between 325 and 1300~MHz of the region associated with IRAS~18162--2048 reveal the presence of the radio inner jet as well as condensations that include HH80-81. The spectral indices of condensations are negative implying dominant non-thermal mechanisms. The central region shows a nearly flat spectral index that we interpret as a combination of thermal emission from the central source and non-thermal contribution from the jet in the vicinity, as our resolution does not permit us to segregate the two. The equipartition magnetic fields estimated using the radio flux densities at 610 MHz are between 116 and 180~$\mu$G. For HH80, where a source with hard X-ray emission has been detected and spectral index suggests a non-thermal process, we consider possibilities that could give rise to this emission from the same population of relativistic electrons that emit in radio. 
We find that a  typical magnetic field value of $\sim3$~$\mu$G for a region of size 0.1~pc, an 
electron number density of $10$~cm$^{-3}$ with volume filling factor of $10^{-3}$, can simultaneously explain the radio and hard X-ray fluxes towards HH80 by the process of inverse-Comptonization of the thermal background infrared photons.
As the magnetic field value is lower than the equipartition value by more than an order of magnitude, the association of the hard X-ray source to HH-80 necessitates further investigation.
The 610~MHz flux densities, measured at two epochs separated by nearly a decade, show variations in flux densities of the condensations that can possibly be linked to the evolution of the knots as well as to the ambient interstellar density.

\vspace*{1cm}

\noindent \textbf{Acknowledgements} \\
We appreciate the inputs from the referee, C. Carrasco-Gonz{\'a}lez, that have helped in improving the presentation of the paper. We thank the staff of the GMRT, who have made the radio observations possible. GMRT is run by the National Centre for Radio Astrophysics of the Tata Institute of Fundamental Research. We are grateful to L. F. Rodr{\'{\i}}guez for providing us with the VLA 20~cm image of this region.

\bibliography{jetref}

\end{document}